\documentclass[a4paper]{article}
\usepackage{graphicx}
\usepackage{amsmath,amssymb}
\usepackage{indentfirst}

\newcommand{\Tr}{{\rm Tr}}

\newcommand{\be}{\begin{equation}}
\newcommand{\ee}{\end{equation}}

\begin{document}

\title{Semiclassical treatment of quantum chaotic transport with a tunnel barrier}
\author{Pedro H. S. Bento, Marcel Novaes \\ Instituto de F\'isica, Universidade Federal de Uberl\^andia\\ Uberl\^andia, MG, 38408-100, Brazil}
\date{}

\maketitle
\begin{abstract}

We consider the problem of a semiclassical description of quantum chaotic transport, when a tunnel barrier is present in one of the leads. Using a semiclassical approach formulated in terms of a matrix model, we obtain transport moments as power series in the reflection probability of the barrier, whose coefficients are rational functions of the number of open channels $M$. Our results are therefore valid in the quantum regime and not only when $M\gg 1$. The expressions we arrive at are not identical with the corresponding predictions from random matrix theory, but are in fact much simpler. Both theories agree as far as we can test.

\end{abstract}

\section{Introduction}

We consider quantum transport through systems with chaotic classical dynamics. A possible setting would be a two dimensional electron gas, shaped as a mesoscopic cavity using semiconductors, connected to source and drain by two attached leads and submitted to a small external voltage \cite{nazarov}. Assuming low temperature and a classical dwell time inside the cavity which is much higher than the Ehrenfest time, the statistical properties of the electronic transport are remarkably universal, i.e. insensitive to system details and depending only on the symmetry class of the problem (for spinless particles, this reduces to presence/absence of time-reversal symmetry) \cite{haake}.

The process of quantum scattering through a cavity with two leads is described by a scattering $S$ matrix, of dimension equal to the number of open channels $M$, which relates incoming to outgoing quantum amplitudes. Conservation of charge implies that $S$ is unitary. Another possibility is to use the eigenvalues of the related transmission matrix $\mathcal{T}$, defined below, which is hermitian. The measurable characteristics of the system, like average and variance of conductance, shot-noise etc., are related to symmetric functions of the eigenvalues of $\mathcal{T}$ \cite{landauer,buttiker}. 

When the corresponding classical dynamics is strongly chaotic and universality is expected, it is fruitful to consider these matrices as random objects, taken from appropriate ensembles. This is the random matrix theory (RMT) approach. We shall focus on systems with broken time-reversal symmetry. When both leads are ideal, i.e. perfectly transmitting, the $S$ is uniformly distributed in the unitary group with Haar measure, and the transmission matrix has a Jacobi distribution \cite{jacobi}. Results valid for a large number of open channels, $M\gg 1$, were obtained along the 1990's \cite{cwj,baranger,jalabert} and were reviewed in \cite{review1997}. Exact results, valid for arbitrary $M$, were produced in the 2000's \cite{savin2,savin3,novaes08,savin4,livan} after a connection with the Selberg integral was exploited \cite{savin1}. 

The semiclassical approximation is a different approach, starting from expressions for matrix elements of quantum observables in terms of the action and stability of classical trajectories \cite{semi1,semi2}. A stationary phase argument establishes that sets of trajectories interfere constructively only if they are correlated, and this correlation is mediated by the existence of close encounters. According to the theory initially developed by Sieber and Richter \cite{sieber1,sieber2} and later further developed by Haake and collaborators \cite{essen3,essen4,essen5} (see also \cite{essen2}), after some integrations over phase space are performed, the calculation of transport moments can be formulated diagrammatically in terms of ribbon graphs, with simple rules determining the contribution of each graph which, when both leads are ideal, are determined by its genus. In that case, it has been established by Berkolaiko and Kuipers \cite{greg1,greg2,greg3} that this approach is equivalent to random matrix theory and provides a microscopic justification for it (the demonstration of this equivalence was vastly simplified by the introduction of semiclassical matrix models in \cite{matrix}). 

In electronic systems, it is more realistic to assume the presence of tunnel barriers in the leads \cite{bar1,bar2,bar3,bar4}, so that channel $i$ has an associated tunnel rate $\Gamma_i$, with $\Gamma_i=1$ being the ideal case. In the random matrix setting this is implemented by introducing the so-called Poisson kernel to model the statistical distribution of the $S$ matrix \cite{poisson1,poisson2}. In the perturbative $M\gg 1$ setting, average and variance of the conductance were obtained in \cite{BB}, while the average shot-noise was considered in \cite{ramos} (see also \cite{ramos2}). Exact formulas for the eigenvalue distribution of $\mathcal{T}$, valid for arbitrary $M$, were derived \cite{kanz1,kanz2} (see also \cite{vidal}) in terms of hypergeometric functions of matrix argument. This theory was then used in \cite{perez} to derive finite-$M$ results for transport moments.

Within the semiclassical theory, modified diagrammatic rules valid in the presence of tunnel barriers \cite{whitney} were able to reproduce the average conductance and shot-noise to leading order in $M^{-1}$, in agreement with RMT. This was later taken further to compute the variance of conductance \cite{jacquod}. These semiclassical investigations have even been capable of taking into account effects that are not captured by random matrix theory, related to the existence of a finite Ehrenfest time (see also \cite{Ehr1,Ehr2,Ehr3,Ehr4}, for example). They have also been modified in order to be applied to the statistics of time delay and to Andreev systems \cite{kuipers,kuipersrichter}.

However, all these previous semiclassical efforts were restricted to the leading orders in $M^{-1}$. It should be possible to push this theory further, since there are still no semiclassical results that are valid in the presence of a tunnel barrier and in the truly quantum regime, i.e. for arbitrary values of $M$. The purpose of the present work is to fill this gap. 

We make use of a novel semiclassical approach which is based on a matrix integral representation \cite{matrix,trs}. The advantage of this method is that all diagrams are built into the theory from the beginning and do not need to be explicitly constructed. This approach has also been used to treat energy-dependent statistics \cite{energy1,energy2}. By appropriately adapting it, we are able to treat the situation with a tunnel barrier. The results we find are in agreement with the corresponding ones obtained within RMT \cite{vidal,perez}, but are in fact much simpler.

\section{Transport Moments}

\subsection{Definitions}

If $X$ is a $N\times N$ matrix with eigenvalues $x_j$, $1\le j\le N$, then
\be p_\lambda(X)=\prod_{i=1}^{\ell(\lambda)}\Tr(X^{\lambda_i})=\prod_{i=1}^{\ell(\lambda)}\sum_{j=1}^N x_j^{\lambda_i}\ee
is the power sum symmetric function, labelled by an integer partition, i.e. a non-decreasing sequence $\lambda=(\lambda_1,\lambda_2,\ldots)$ of $\ell(\lambda)$ positive integers. If $\sum_i \lambda_i=n$ we say $\lambda$ partitions $n$ and denote this by $\lambda\vdash n$ or $|\lambda|=n$. If $\pi$ is some permutation with cycle type $\lambda$, then this function can also be written as
\be p_\lambda(X)=\sum_{i_1,...i_n}\prod_{k=1}^n X_{i_{\pi(k)},i_k}.\ee

We assume a chaotic cavity with two leads, supporting $N_1$ and $N_2$ open channels. The total number of channels is 
\be M=N_1+N_2.\ee The $S$-matrix is given by $S=\begin{pmatrix} r&t\\t'&r'\end{pmatrix}$, where $r$ is a $N_1\times N_1$ reflection block and $t$ is a $N_2\times N_1$ transmission block (and similarly for $r'$ and $t'$). The $N_1\times N_1$ transmission matrix is $\mathcal{T}=t^\dagger t$. The dimensionless transport moments are the functions $p_\lambda(\mathcal{T})$: for instance, the conductance is $p_{1}(\mathcal{T})$, while the shot-noise is $p_{1}(\mathcal{T})-p_{2}(\mathcal{T})$. These moments are related to the statistical properties of the electric current in the system as a function of time: conductance and shot-noise, for instance, are related to average and variance of current. 

In a realistic system these transport moments are wildly fluctuating functions of the energy and therefore have a random behaviour of their own, and we can talk about their statistical properties. An ensemble average (in random matrix theory) or a local energy average (in semiclassical theory) may be introduced. We denote both these averages by $\langle p_\lambda(\mathcal{T})\rangle$. The variance of conductance, for example, would be related to $\langle p_1^2\rangle-\langle p_1\rangle^2$. Notice that $p_1^2=p_{1,1}$. Transport moments associated with partitions with more than one part are sometimes called `nonlinear statistics' (but we avoid this terminology).

Transport moments can also be encoded in a different family of symmetric functions called Schur functions. These are given by
\be s_\lambda(X)=\frac{\det\left(x_i^{n+\lambda_i-i}\right)}{\det\left(x_i^{n-i}\right)}=\frac{\det\left(x_i^{n+\lambda_i-i}\right)}{\Delta(X)},\ee
where $n=|\lambda|$ and 
\be \Delta(X)=\prod_{i=1}^N\prod_{j=i+1}^N (x_j-x_i)\ee
is called the Vandermonde of $X$. The set of Schur functions $\{s_\lambda,\lambda\vdash n\}$ spans the vector space of homogeneous symmetric polynomials of degree $n$. They are related to power sums by
\be p_\mu(X)=\sum_{\lambda\vdash |\mu|}\chi_\lambda(\mu)s_\lambda(X),\ee
where $\chi_\lambda(\mu)$ are the irreducible characters of the permutation group.

An important role is played by the value of the Schur function when all its arguments are equal to $1$. In that case 
\be s_\lambda(1^N)=\frac{d_\lambda}{n!}[N]_\lambda,\ee where $d_\lambda=\chi_\lambda(1^n)$ is the number of standard Young tableaux of shape $\lambda$ and $[N]_\lambda$ is a generalization of the rising factorial given by
\be [N]_\lambda=\prod_{i=1}^{\ell(\lambda)}\frac{(N+\lambda_i-i)!}{(N-i)!}.\ee This can also be written as a product over the contents of the Young diagram (see Appendix A), \be\label{Nc} [N]_\lambda=\prod_{(i,j)\in\lambda}(N+j-i).\ee

Finally, let us mention that the product of two such functions can be written as a linear combination of them according to 
\be s_\lambda(X)s_\mu(X)=\sum_\nu C^\nu_{\lambda\mu}s_\nu(X).\ee The quantities $C^\nu_{\lambda\mu}$ are called Littlewood-Richardson coefficients. They are different from zero only if $|\nu|= |\lambda|+|\mu|$ and additionally $\nu$ contains both $\lambda$ and $\mu$. We say $\nu$ contains $\mu$, $\nu\supset \mu$, when $\nu_i\ge \mu_i$ for all $i$.

\subsection{Random matrix theory}

In the ideal case when there are no tunnel barriers, the joint probability distribution of the $N_1$ eigenvalues of $\mathcal{T}$ is, assuming without loss of generality that $N_1\le N_2$, given by
\be P(T)=\frac{1}{\mathcal{Z}_0} |\Delta(T)|^2\prod_{i=1}^{N_1}T_i^{N_2-N_1},\ee where $\mathcal{Z}_0$ is a normalization constant.
The calculation of the average value of any Schur function amounts to a Selberg-like integral \cite{kaneko,kadell},
\be\label{idealrmt} \langle s_\mu(T)\rangle=\int_{[0,1]^{N_1}} s_\mu(T)P(T)dT=\frac{d_\mu}{n!}\frac{[N_1]_\mu[N_2]_\mu}{[M]_\mu}.\ee

In the non-ideal situation, Vidal and Kanzieper obtained the joint probability distribution of reflection eigenvalues $R_i=1-T_i$, assuming only one of the leads contains a tunnel barrier and time-reversal symmetry is broken. Assuming the $N_2$ channels in the second lead are ideal, while in the first lead the tunnelling probabilities $\Gamma_i$ of each channel are collected in the matrix $\Gamma$, their result is that 
\be P(R)\propto\det(\Gamma)^M\det(\mathcal{F})\frac{\Delta(R)}{\Delta(\Gamma)}\prod_{i=1}^{N_1}(1-R_i)^{N_2-N_1},\ee
where $\mathcal{F}$ is a matrix with elements given in terms of a hypergeometric function,
\be \mathcal{F}_{ij}={}_2F_1(N_2+1,N_2+1;1;(1-\Gamma_i)R_j).\ee

The above result was used in \cite{perez} to obtain average transport moments. In terms of Schur functions of reflection eigenvalues, it was shown that\label{smu}
\be \langle s_\lambda(R)\rangle=\det(\Gamma)^M\sum_{\rho}s_\rho(1-\Gamma)\frac{[M]_\rho^2}{[N_1]_\rho^2}\sum_{\nu}C^\nu_{\lambda\rho}\frac{d_\nu}{|\nu|!}\frac{[N_1]_\nu^2}{[M]_\nu},\ee where the infinite sum over $\rho$ includes all possible partitions and the quantities $C^\nu_{\lambda\rho}$ are the Littlewood-Richardson coefficients. In the regime of weakly non-ideal leads, $\Gamma_i\approx 1$, this result can be seen as a perturbative expansion in the small variable $1-\Gamma$.

In this work we shall further assume that in the first lead all tunnelling probabilities are equal, and we will express them in terms of an opacity parameter $\gamma$, which is the reflection probability of the barrier, $\Gamma_i=1-\gamma$. Then we can use the relation
\be s_\rho(\gamma)=\gamma^{|\rho|}\frac{d_\rho[N_1]_\rho}{|\rho|!}\ee to write
\be\label{rmtfinal} \langle s_\lambda(R)\rangle=(1-\gamma)^{N_1M}\sum_{\rho}\gamma^{|\rho|}\frac{d_\rho}{|\rho|!}\frac{[M]_\rho^2}{[N_1]_\rho}\sum_{\nu}C^\nu_{\lambda\rho}\frac{d_\nu}{|\nu|!}\frac{[N_1]_\nu^2}{[M]_\nu}.\ee

The average value of a Schur function of the transmission eigenvalues can be obtained from the above equation by using the binomial-like theorem \cite{macdonald}
\be\label{binomial} s_\mu(T)=s_\mu(1-R)=\sum_{\lambda\subset\mu} (-1)^{|\lambda|}B_{\mu,\lambda}(N_1)s_\lambda(R),\ee in which 
\be B_{\mu,\lambda}(N_1)=\det\left({N_1+\mu_i-i \choose N_1+\lambda_j-j}\right)=\frac{[N_1]_\mu}{[N_1]_\lambda}\frac{d_{\mu/\lambda}}{(|\mu|-|\lambda|)!},\ee
with
\be d_{\mu/\lambda}=(|\mu|-|\lambda|)!\det\left(\frac{1}{(\mu_i-i-\lambda_j+j)!}\right)\ee
being the number of standard Young tableaux of shape $\mu/\lambda$. 

This leads to a rather cumbersome final expression for $\langle s_\mu(T)\rangle$, that we have not been able to simplify. Our initial aim was to reproduce this result using the semiclassical approximation. However we shall see that in fact that theory leads to a much cleaner formula. 

\section{The semiclassical approximation}

\subsection{Action correlations}

The leads that connect the chaotic cavity to the outside world have widths $W_1$ and $W_2$. These must be classically small in order to ensure that the dwell time $\tau_D$ be long enough for the dynamics to be strongly chaotic. On the other hand, in the semiclassical regime of small $\hbar$ the number of open channels $N_i\sim W_i/\hbar$ will be large (the total energy of the electrons is fixed). 

The semiclassical approximation starts by writing the elements of the $S$ matrix as sums over classical trajectories:
\be S_{oi}=\frac{1}{\sqrt{M\tau_D}}\sum_{\alpha:i\to o}A_\alpha e^{i\mathcal{S}_\alpha/\hbar},\ee
where each trajectory $\alpha$ starts at channel $i$ and ends at channel $o$, having action $\mathcal{S}_\alpha$ (the prefactor $A_\alpha$ is related to
the trajectory's stability). 

The calculation of a transport moment like
\be p_\lambda(\mathcal{T})=\sum_{\vec{i}=1}^{N_1}\prod_{k=1}^n (t^\dagger t)_{i_{\pi(k)},i_k}=\sum_{\vec{i}=1}^{N_1}\sum_{\vec{o}=1}^{N_2}\prod_{k=1}^n t^\dagger_{i_{\pi(k)},o_k} t_{o_{k},i_k},\ee
with $\pi$ being any permutation that has cycle type $\lambda$, requires multiple
sums over trajectories, \be\label{multiple}
p_\lambda(\mathcal{T})=\frac{1}{M^n\tau_D^n}\prod_{k=1}^n\sum_{i_k,o_k}\sum_{\alpha_k,\sigma_k}A_\alpha
A^*_\sigma e^{i(\mathcal{S}_\alpha-\mathcal{S}_\sigma)/\hbar},\ee with the understanding that $\alpha_k$ goes from $i_k$ to $o_k$, while $\sigma_k$ goes from $i_{\pi(k)}$ to $o_k$. The quantity $A_\alpha=\prod_k A_{\alpha_k}$ is a collective stability, while $\mathcal{S}_\alpha=\sum_k
\mathcal{S}_{\alpha_k}$ is the collective action of the $\alpha$ trajectories, and analogously for $\sigma$.

The transport moment (\ref{multiple}) is a strongly fluctuating function of the energy. Its local average value can be computed under a stationary phase
approximation, which leads to the condition that the set of $\alpha$ trajectories has almost the same collective action as the set of $\sigma$ trajectories. These so-called action correlations exist when the $\alpha$'s and $\sigma$'s are piecewise almost equal, except in small regions where an encounter takes place. A $q$-encounter is a region where $q$ pieces of trajectories run nearly parallel and the $\sigma$'s are permuted with respect to the  $\alpha$'s (we are considering only systems with broken time-reversal symmetry, so $\sigma$ trajectories never run opposite to $\alpha$ trajectories). 

\begin{figure}[b]
\includegraphics[scale=1.3,clip]{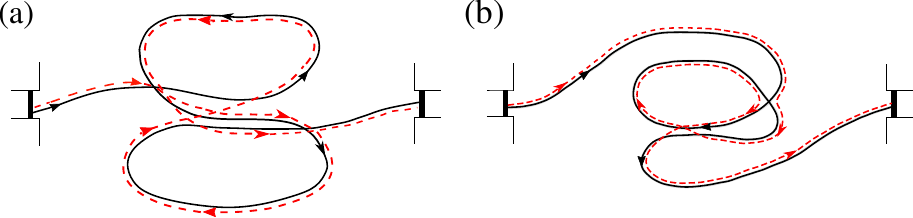}
\caption{Action correlated trajectories (in solid and dashed lines) that contribute to the semiclassical evaluation of the conductance. a) Trajectories differ by a $3$-encounter; b) Trajectories differ by two $2$-encounters. Black rectangles represent the leads.} 
\end{figure}

As illustration, we present in Figure 1 two contributions to the simplest transport moment, the average conductance $\langle p_1(\mathcal{T})\rangle$. Trajectory $\alpha$ is depicted in solid line, while $\sigma$ is in dashed line. In panel a) we have a $3$-encounter, while in panel b) we have two $2$-encounters. For the sake of visual clarity, the encounters are greatly magnified so that their internal structure is visible. Also we do not try to reproduce the actual trajectories which would be extremely convoluted and chaotic. For more details regarding this theory, and plenty more figures, we refer the reader to previous works.

When a tunnel barrier is present, say in the left lead for instance, action correlations may be of a slightly different nature, as trajectories that hit the barrier from the inside may fail to tunnel out and, instead, may be reflected back into the cavity \cite{whitney,kuipers}. When this happens a trajectory will be composed of two or more parts, corresponding to its excursions between hits in the barrier. Two trajectories may then differ in the order of these excursions, while still having the same action. This is illustrated in Figure 2. In panel a) $\alpha$ and $\sigma$ hit the left lead twice before leaving the cavity; they differ in order they traverse those two `loops'. In panel b) $\alpha$ and $\sigma$ hit the left lead once and, in addition, there is a $2$-encounter. 

These special situations may be interpreted in terms of `encounters in the lead'. Fig.2a is then viewed as a degenerate case of Fig.1a, in which the $3$-encounter happens in the lead, namely, it is replaced by reflections. Analogously, Fig.2b is viewed as a degenerate case of Fig.1b, in which one of the $2$-encounters happens in the lead while the other one remains inside the cavity.

\begin{figure}[t]
\includegraphics[scale=1.3,clip]{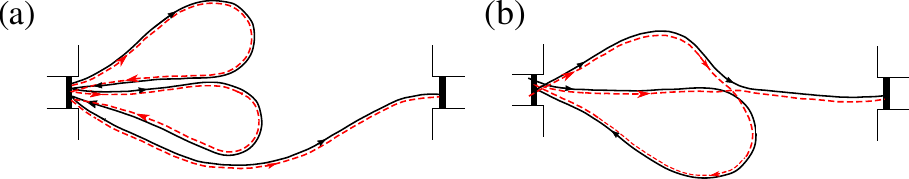}
\caption{Action correlated trajectories (in solid and dashed lines) that contribute to the semiclassical evaluation of the conductance when a tunnel barrier is present in the left lead. They may be seen as degenerate cases of the ones in Figure 1, in which the encounter takes place in the lead. Black rectangles represent the leads.} 
\end{figure}

\subsection{Semiclassical diagrams}

Sets of action-correlated trajectories can be represented by diagrams which are ribbon graphs. A $q$-encounter becomes a vertex of valence $2q$. 
The pieces of trajectories between vertices become oriented ribbons, bordered by one of the $\alpha$ trajectories on one side and one of the $\sigma$ trajectories on the other. We show in Figure 3 the ribbon graphs corresponding to the trajectories shown in Figure 1. 

\begin{figure}[b]
\includegraphics[scale=0.75,clip]{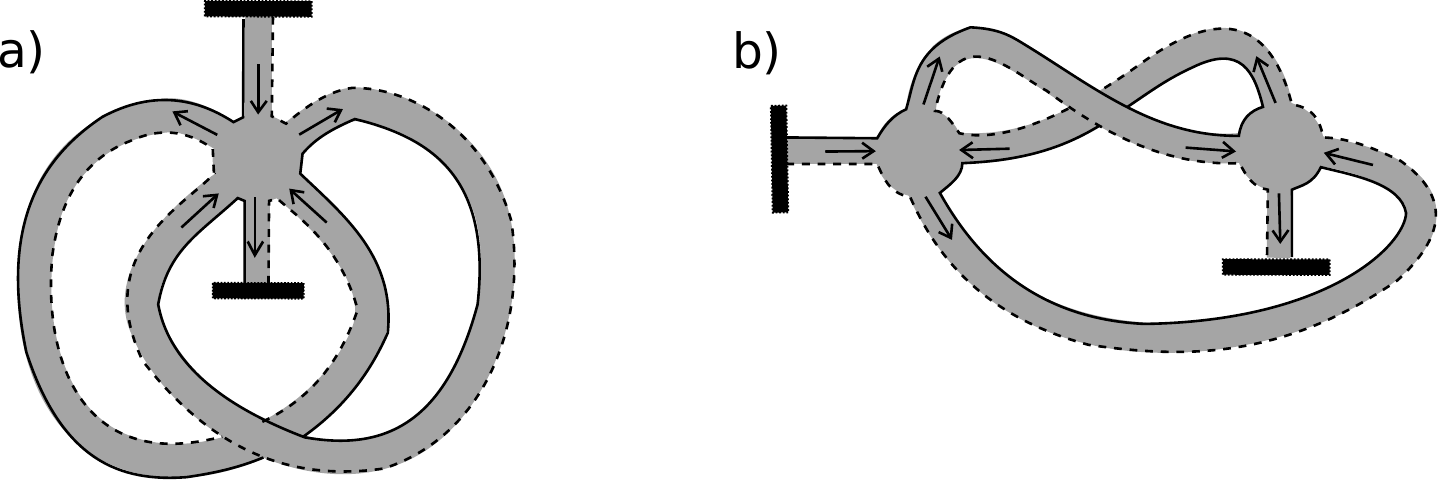}
\caption{Diagrams that represent the trajectories in Fig.1. Encounters are depicted as vertices, arcs of trajectories are depicted as ribbons, bordered by an $\alpha$ trajectory on one side (solid line) and a $\sigma$ trajectory on the other (dashed line). Black rectangles represent the leads.} 
\end{figure}

In the ideal case when there are no tunnel barriers, after the appropriate phase-space integrals are performed the semiclassical theory boils down to summing over diagrams, with diagrammatic rules that are as follows: the contribution of a diagram is multiplied by
\begin{itemize}
\item $M^{-1}$ for each ribbon,
\item $-M$ for each vertex,
\item $N_1$ for each channel where a trajectory begins,
\item $N_2$ for each channel where a trajectory ends.
\end{itemize}

For example, the leading contribution to the conductance comes from the trivial diagram with no encounters and identical trajectories, $\alpha=\sigma$. This gives $N_1N_2/M$. The next contributions are sketched in Figure 3. Panel a) has four ribbons and one $3$-vertex, giving $-N_1N_2/M^3$; Panel b) has five ribbons and two $2$-vertices, giving $N_1N_2/M^3$.

As discussed in \cite{whitney,kuipers}, in the non-ideal case when tunnel barriers are present, the semiclassical diagrammatic rules must be modified. Assuming that the second lead is ideal and that in the first lead all tunnelling probabilities are equal, $\Gamma_i=1-\gamma$, the contributions become
\begin{itemize}
\item $(N_1(1-\gamma)+ N_2)^{-1}=(M-N_1\gamma)^{-1} $ for each ribbon,
\item $-N_1(1-\gamma^q)-N_2=-M+N_1\gamma^q$ for each vertex of valence $2q$,
\item $N_1(1-\gamma)$ for each channel where a trajectory begins,
\item $N_2$ for each channel where a trajectory ends.
\item $\gamma$ for each encounter happening at the first lead.
\end{itemize}

The leading order contribution to the calculation of the conductance, for example, becomes $N_1N_2(1-\gamma)/(M-N_1\gamma)$. The diagrams in Fig.3a) and Fig.3b), on the other hand, now give 
\be \frac{N_1N_2(1-\gamma)(-M+N_1\gamma^3)}{(M-N_1\gamma)^4}\quad \text{and} \quad \frac{N_1N_2(1-\gamma)(-M+N_1\gamma^2)^2}{(M-N_1\gamma)^5},\ee respectively. Moreover, we must now allow encounters in the lead (these do not count as vertices in the diagrammatic theory, however). The trajectories in Figure 2 and their diagrams in Figure 4 thus come into play. Fig.4a has three ribbons, no vertices and two reflections; Fig.4b has four ribbons, one $2$-vertex and one reflection. Their contributions are 
\be \frac{N_1N_2(1-\gamma)\gamma^2}{(M-N_1\gamma)^3}\quad \text{and} \quad \frac{N_1N_2(1-\gamma)\gamma(-M+N_1\gamma^2)}{(M-N_1\gamma)^4},\ee respectively.

\begin{figure}[b]
\includegraphics[scale=0.75,clip]{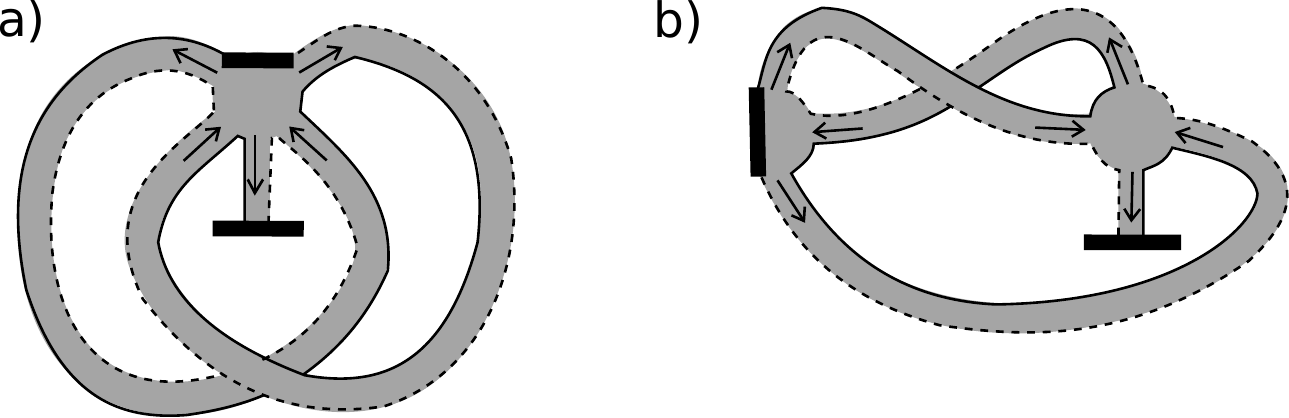}
\caption{Diagrams that represent the trajectories in Fig.2. Encounters may now happen `in' the first lead, as trajectories may be reflected by the tunnel barrier back inside the cavity. Black rectangles represent the leads.} 
\end{figure}

\section{Matrix model with a tunnel barrier}

\subsection{Ideal case}

In the ideal case when there are no tunnel barriers, the diagrammatic rules can be implemented by means of the matrix integral 
\be \label{matrix}\left\langle \prod_{k=1}^n t^\dagger_{i_{\pi(k)},o_k}t_{o_k,i_k} \right\rangle=\lim_{N\to 0}\frac{1}{\mathcal{Z}}\int e^{-M\sum_{q=1}^\infty \frac{1}{q}\Tr(Z^\dag Z)^q} \prod_{k=1}^n Z^\dagger_{i_{\pi(k)},o_k}Z_{o_k,i_k}dZ,\ee
where we integrate over $N\times N$ complex matrices $Z$. The quantity
\be \mathcal{Z}=\int e^{-M\Tr(ZZ^\dag)}dZ=\frac{N!}{M^{N^2}}\prod_{j=1}^{N-1}j!^2\ee
is a normalization constant. 

The diagrammatic approach to (\ref{matrix}) proceeds from keeping $e^{-M\Tr(Z^\dag Z)}$ as a Gaussian measure and expanding the remaining exponentials as power series. The integral is then performed using the well known Wick's rule \cite{matrix,energy1,morris,francesco}. The product of $Z^\dagger$ and $Z$ matrix elements represent the channels. The diagrams thus produced are exactly like the semiclassical ribbon graphs, with the same diagrammatic rules, $M^{-1}$ for each ribbon, $-M$ for each vertex ($N_1$ and $N_2$ do not appear yet since we are keeping fixed the indices $i$ and $o$). 

However, when producing all possible connections as per Wick's rule, summation over free indices in the traces will produce powers in the dimension $N$. These are closed cycles that correspond to periodic orbits. Since the semiclassical sum does not include such orbits, we let $N\to 0$ at the end of the calculation to get rid of them. 

The integral (\ref{matrix}) was computed exactly in \cite{matrix} using singular value decomposition. Here we briefly present a different method. First, we sum the exponential and sum over $i$ from $1$ to $N_1$ and over $o$ from $1$ to $N_2$, and write it in the form
\be\label{easy} \langle p_\lambda(T) \rangle=\lim_{N\to 0}\frac{1}{\mathcal{Z}}\int \det(1-Z^\dag Z)^M p_\lambda(Z^\dagger Q_2ZQ_1)dZ,\ee
where $\lambda$ is the cycle type of $\pi$ and 
\be Q_1=\begin{pmatrix} 1_{N_1} & 0\\0 & 0_{N_2}\end{pmatrix},\quad Q_2=\begin{pmatrix} 0_{N_1} & 0\\0 & 1_{N_2}\end{pmatrix}\ee
are projectors, with $1_N$ and $0_N$ being the identity and the null matrix in dimension $N$. Then, expand the power sum into Schur functions and use the fact that (see equation (18) in \cite{fyodorov})
\be \int \det(1-Z^\dag Z)^{M-2N}s_\mu(Z^\dagger Q_2ZQ_1)dZ=\frac{s_\mu(Q_1)s_\mu(Q_2)}{s_\mu(1^M)}.\ee
Finally, since $s_\mu(Q_i)=s_\mu(1^{N_i})=\frac{d_\mu}{n!}[N_i]_\mu$, we obtain 
\be \langle p_\lambda(T) \rangle=\frac{1}{n!}\sum_{\mu\vdash n}\chi_\mu(\lambda)d_\mu\frac{[N_1]_\mu[N_2]_\mu}{[M]_\mu}.\ee
The above calculation is the semiclassical derivation of transport moments in the ideal case, which agrees exactly with RMT (\ref{idealrmt}).

\subsection{Including the tunnel barrier}

In the non-ideal case when a tunnel barrier is present in the first lead, we have different diagrammatic rules to implement. If there were no encounters in the lead, we could propose the modified matrix model
\be \lim_{N\to 0}\frac{(1-\gamma)^{n}}{\mathcal{Z}}\int \exp\left[-\sum_{q=1}^\infty\frac{(M-N_1 \gamma^q)}{q}\Tr(Z^\dag Z)^q\right] \prod_{k=1}^n Z^\dagger_{o_{\pi(k)},i_k}Z_{o_k,i_k}dZ,\ee
where 
\be \mathcal{Z}=\int e^{-(M-N_1\gamma)\Tr(ZZ^\dag)}dZ=\frac{N!}{(M-N_1\gamma)^{N^2}}\prod_{j=1}^{N-1}j!^2.\ee
The prefactor $(1-\gamma)^{n}$ corresponds to all the trajectories entering the cavity through the barrier. The Gaussian measure $e^{-(M-N_1\gamma)\Tr (Z^\dag Z)}$ produces $(M-N_1\gamma)^{-1}$ for each ribbon; expanding the rest of the exponential produces $-M+N_1\gamma^{q}$ for each $q$-encounter, so the diagrammatic rules would be indeed correct. 

However, we must also incorporate encounters in the first lead. Information about what happens at the leads must be contained in the factor $\prod_{k=1}^n Z^\dagger_{o_{\pi(k)},i_k}Z_{o_k,i_k}$, because diagrammatically each matrix element from $Z$ is a pair of trajectories entering the cavity, while each matrix element from $Z^\dagger$ is a pair of trajectories leaving the cavity. So we must introduce some modification to this term.

When an encounter happens in the lead, a vertex-like structure is produced. In Fig.4a this pseudo-vertex has valence 5, while in Fig.4b it has valence 3. It is as if a matrix element from $Z$ got replaced by a matrix element from $(ZZ^\dag ZZ^\dag Z)$ in the first case and $(ZZ^\dag Z)$ in the second. To produce a pseudo-vertex of valence $2m+1$ we should replace $Z$ by $Z(Z^\dag Z)^m$. This must be accompanied by $\gamma^m$ according to the diagrammatic rules. Encounters in the lead can thus be implemented by means of a geometric series, and we therefore postulate the integral
\be\label{comp} \langle p_\lambda(T)\rangle=\lim_{N\to 0}\frac{1}{\mathcal{Z}}\int e^{\left[-\sum_{q=1}^\infty\frac{(M-N_1 \gamma^q)}{q}\Tr(Z^\dag Z)^q\right]} \prod_{k=1}^n Z^\dagger_{i_{\pi(k)},o_k}\left(Z\frac{1}{1-\gamma Z^\dag Z}\right)_{o_k,i_k}dZ.\ee

\subsection{Exact calculation}

The integral in (\ref{comp}) is more complicated than the one in (\ref{matrix}). In order to compute it, we introduce the singular value decomposition $Z=UDV$ and perform first the integration of the angular variables $U$ and $V$ over the unitary group $\mathcal{U}(N)$. This is done in \cite{matrix} and the result is
that the integral \be \sum_{\vec{i}=1}^{N_1}\sum_{\vec{o}=1}^{N_2}\int dUdV (V^\dag D U^\dag)_{i_{\pi(k)},o_k}\left(U D V\frac{1}{1-\gamma X}\right)_{o_k,i_k}\ee equals \be\sum_{\mu\vdash n}\frac{[N_1]_\mu[N_2]_\mu}{[N]_\mu^2}\chi_\mu(\lambda)s_\mu\left(\frac{X}{1-\gamma X}\right),\ee where $X=D^2$.

Let us mention in passing that the function $s_\mu\left(\frac{X}{1-\gamma X}\right)$ is a particular case of the canonical stable Grothendieck functions studied in \cite{yeli}.

We now turn to the radial integral, i.e. the integral over the diagonal matrix $X$. Summing the series in the exponent we find that this is
\be \int dX|\Delta(X)|^2\det(1-X)^M\det(1-\gamma X)^{-N_1}s_\mu\left(\frac{X}{1-\gamma X}\right),\ee where $|\Delta(X)|^2$ is the Jacobian of the singular value decomposition. In order to make progress, we must consider this integral in the form of a power series in $\gamma$. To express the second determinant, we resort to the well known Cauchy identity,
\be \det(1-\gamma X)^{-N_1}=\sum_\omega s_\omega(\gamma)s_\omega(X),\ee
where the infinite sum is over all possible partitions and the first Schur function has $N_1$ variables equal to $\gamma$. On the other hand, the Schur function with the awkward argument can be written as
\be\label{awk} s_\mu\left(\frac{X}{1-\gamma X}\right)=\sum_{\rho\supset \mu} \gamma^{|\rho|-|\mu|}A_{\mu\rho}s_\rho(X),\ee
with the coefficients being given in terms of a determinant with binomial elements:
\be\label{A} A_{\mu\rho}=\det\left({\rho_i-i \choose \mu_j-j}\right).\ee We derive expansion (\ref{awk}) in Appendix B.

Littlewood-Richardson coefficients allow us to write 
\be\label{LR} s_\omega(X)s_\rho(X)=\sum_{\nu} C^\nu_{\omega,\rho}s_\nu(X)\ee and arrive at a well known Selberg-like integral \cite{kaneko,kadell},
\be \frac{1}{\mathcal{Z}}\int \det(1-X)^Ms_\nu(X)|\Delta(X)|^2dX=(M-\gamma N_1)^{N^2}\frac{d_\nu[N]_\nu^2}{|\nu|![M]_\nu}\prod_{j=0}^{N-1}\frac{(M+j)!}{(M+N+j)!}.\ee

Having computed all the integrals, this is the time to consider the limit $N\to 0$. First, $(M-\gamma N_1)^{N^2}\to 1$. Also,
\be \prod_{j=0}^{N-1}\frac{(M+j)!}{(M+N+j)!}\to \prod_{j=0}^{N-1}\frac{(M+j)!}{(M+j)!}\to 1.\ee Finally, we need to deal with 
\be\label{limit} \lim_{N\to 0}\frac{[N]_\nu}{[N]_\mu}.\ee From the expression of $[N]_\nu$ in terms of contents, Eq. ({\ref{Nc}), we know that, for small $N$,
\be [N]_\nu= t_\nu N^{D(\nu)}+O(N^{D(\nu)+1}),\ee where $t_\nu$ is the product of all non-zero contents,
\be t_\nu=\prod_{\substack{(i,j)\in \nu \\ i\neq j}}(j-i)\ee and $D(\nu)$ is the size of the Durfee square of $\nu$, i.e. the side length of the largest square diagram contained in $\nu$. Since $\nu\supset \rho$ because of (\ref{LR}) and $\rho\supset \mu$ because of (\ref{A}), we have $D(\nu)\ge D(\mu)$, so the limit (\ref{limit}) exists and is different from zero only if $D(\nu)=D(\mu)$. In this case it equals $t_\nu/t_\mu$, which we can also write as $t_{\nu/\mu}$ (see Appendix A for more details).

Collecting all the terms, what we have for $\langle p_\lambda(T)\rangle$  is
\be\label{compl} (1-\gamma)^{n}\sum_{\mu\vdash n}[N_1]_\mu[N_2]_\mu\chi_\mu(\lambda)\sum_\omega s_\omega(\gamma)\sum_{\rho\supset \mu} \gamma^{|\rho|-n}A_{\mu\rho}\sum_{\substack{\nu\supset \mu\\D(\nu)=D(\mu)}} C^\nu_{\omega,\rho}\frac{d_\nu}{|\nu|!}\frac{1}{[M]_\nu}t_{\nu/\mu}^2.\ee

\subsection{Simplification}

Expression (\ref{compl}) can be simplified if we notice that  
\be \sum_\omega s_\omega(\gamma)C^\nu_{\omega,\rho}=s_{\nu/\rho}(\gamma)=\gamma^{|\nu|-|\rho|}s_{\nu/\rho}(1^{N_1})\ee is a skew-Schur function. From the Jacobi-Trudi expression in terms of complete symmetric functions we can see that this is \be s_{\nu/\rho}(1^{N_1})=\det\left({N_1+\nu_i-i-\rho_j+j-1 \choose \nu_i-i-\rho_j+j}\right). \ee Then, we get
\be\label{p1} \langle p_\lambda(T)\rangle=(1-\gamma)^{n}\sum_{\mu\vdash n}[N_1]_\mu[N_2]_\mu\chi_\mu(\lambda)\sum_{\substack{\nu\supset \mu\\D(\nu)=D(\mu)}} \gamma^{|\nu|-n}E_{\mu\nu}(N_1)\frac{d_\nu}{|\nu|!}\frac{1}{[M]_\nu}t_{\nu/\mu}^2,\ee where 
\be E_{\mu\nu}(N_1)=\sum_{\mu\subset\rho\subset\nu} A_{\mu\rho}s_{\nu/\rho}(1^{N_1}).\ee 

The calculation of $E_{\mu\nu}$ is possible by using Lemma 9.1 from \cite{yeli}, a version of the Cauchy-Binet formula which states that, if $H_{\nu\mu}=\sum_\rho F_{\nu\rho}G_{\rho\mu}$ with $F_{\nu\rho}=\det(f_{\nu_i-i,\rho_j-j})$ and $G_{\rho\mu}=\det(g_{\rho_i-i,\mu_j-j})$, then 
\be H_{\nu\mu} =\det\left(\sum_k f_{\nu_i-i,k}g_{k,\mu_j-j}\right).\ee
Applied to our problem, this Lemma gives
\begin{align} E_{\mu\nu}(N_1)&=\det\left(\sum_k {N_1+\nu_i-i-k-1 \choose \nu_i-i-k}{k \choose \mu_j-j}\right),\\&=\det\left({N_1+\nu_i-i\choose N_1+\mu_j-j}\right)=\frac{[N_1]_\nu}{[N_1]_\mu}\frac{d_{\nu/\mu}}{(|\nu|-n)!}.\end{align}

If, instead of computing $\langle p_\lambda(T)\rangle$, we choose to write the average value of a Schur function, then for $\mu\vdash n$ we get
\be\label{semires} \langle s_\mu(T)\rangle=(1-\gamma)^{n}[N_2]_\mu\sum_{m=0}^\infty \frac{\gamma^m}{m!}\sum_{\substack{\nu\vdash n+m\\\nu\supset\mu\\D(\nu)=D(\mu)}} \frac{[N_1]_\nu}{[M]_\nu}\frac{d_\nu d_{\nu/\mu}}{|\nu|!}t_{\nu/\mu}^2.\ee The $m=0$ term is given by $\nu=\mu$ and indeed coincides with the result for the ideal case. 

The semiclassical result in (\ref{semires}) is actually much simpler than the cumbersome expression from random matrix theory, obtained by combining (\ref{rmtfinal}) and (\ref{binomial}). That these two results, derived from different theories using different methods, are in fact identical is not obvious at all, but we have checked that this is indeed true for all partitions up to $n=5$ and all orders in $\gamma$ up to 6. 

In particular, we can write a rather simple formula for the average conductance. When $\mu=1$ we have that $\nu$ must be a hook, $\nu=(m+1-k,1^k)$, and \be d_{\nu/1}=d_{\nu}={m\choose k}.\ee The total content is $t_{\nu/1}=(m-k)!k!$, so that
\be \langle s_1(T)\rangle=(1-\gamma)\frac{N_1N_2}{M}\sum_{m=0}^\infty \frac{\gamma^m}{m+1}\sum_{k=0}^{m} \frac{(N_1+1)^{m-k}(N_1-1)_k}{(M+1)^{m-k}(M-1)_k},\ee
where $(M)^k$ and $(M)_k$ are the usual rising and falling factorials.

\section{Conclusion}

By using a formulation in terms of matrix integrals, we developed a semiclassical approach to quantum chaotic transport that is able to describe systems with a tunnel barrier in one of the leads. Our results incorporate the barrier in a perturbative way, but are exact in the number of channels, i.e. there is no large-$M$ expansion. 

Exact agreement was found, as far as can be computed, with corresponding results from random matrix theory. However, it came as a surprise that the semiclassical expression for transport moments is actually much simpler than the RMT one, which is not very explicit as it depends on Littlewood-Richardson coefficients. In particular, we found a nice semiclassical formula for the average conductance.

Let us mention that, by adapting the diagrammatic rules, it would also be possible to incorporte an energy dependence into the problem and compute the average value of quantities containing the $S$ matrix at energy $E$ and its adjoint $S^\dagger$ at energy $E+\epsilon$. All that is required is to replace $M-N_1\gamma^q$ by $M(1-iq\epsilon)-N_1\gamma^q$ in the exponent of (\ref{comp}) and the calculation would proceed similarly. 

Another extension of the present work could be the treatment of time-reversal invariant systems. As discussed in \cite{trs,energy2} the matrix model approach can be used in that case by replacing complex matrices with real ones and Schur polynomials with zonal polynomials. That topic deserves further exploration. 

\section*{Acknowledgments}
We thank Jack Kuipers for helping us understand diagrams with encounters in the lead. We thank user61318 of MathOverflow for bringing reference \cite{yeli} to our attention in their answer to question 364518. Financial support from CAPES and from CNPq, grant 306765/2018-7, are gratefully acknowledged.

\section*{Appendix A}

A partition $\lambda\vdash n$ can be represented by a diagram, which is a left-justified collection of boxes containing $\lambda_i$ boxes in line $i$. In Figure 5 we show the diagram associated with $\lambda=(4,4,2,2,1)$. The content of the $j$th box in line $i$ is given by $j-i$. These contents are also shown in Figure 5. The Durfee square is highlighted in grey, its diagonal contains the boxes with zero content. In this case we have
\be [N]_{(4,4,2,2,1)}=(N^2-9)(N^2-4)^2(N^2-1)^2(N+4)N^2=576N^2+O(N^3).\ee Notice that the product of all non-zero contents is $t_\lambda=576$.

\begin{figure}[t]
\includegraphics[scale=0.35,clip]{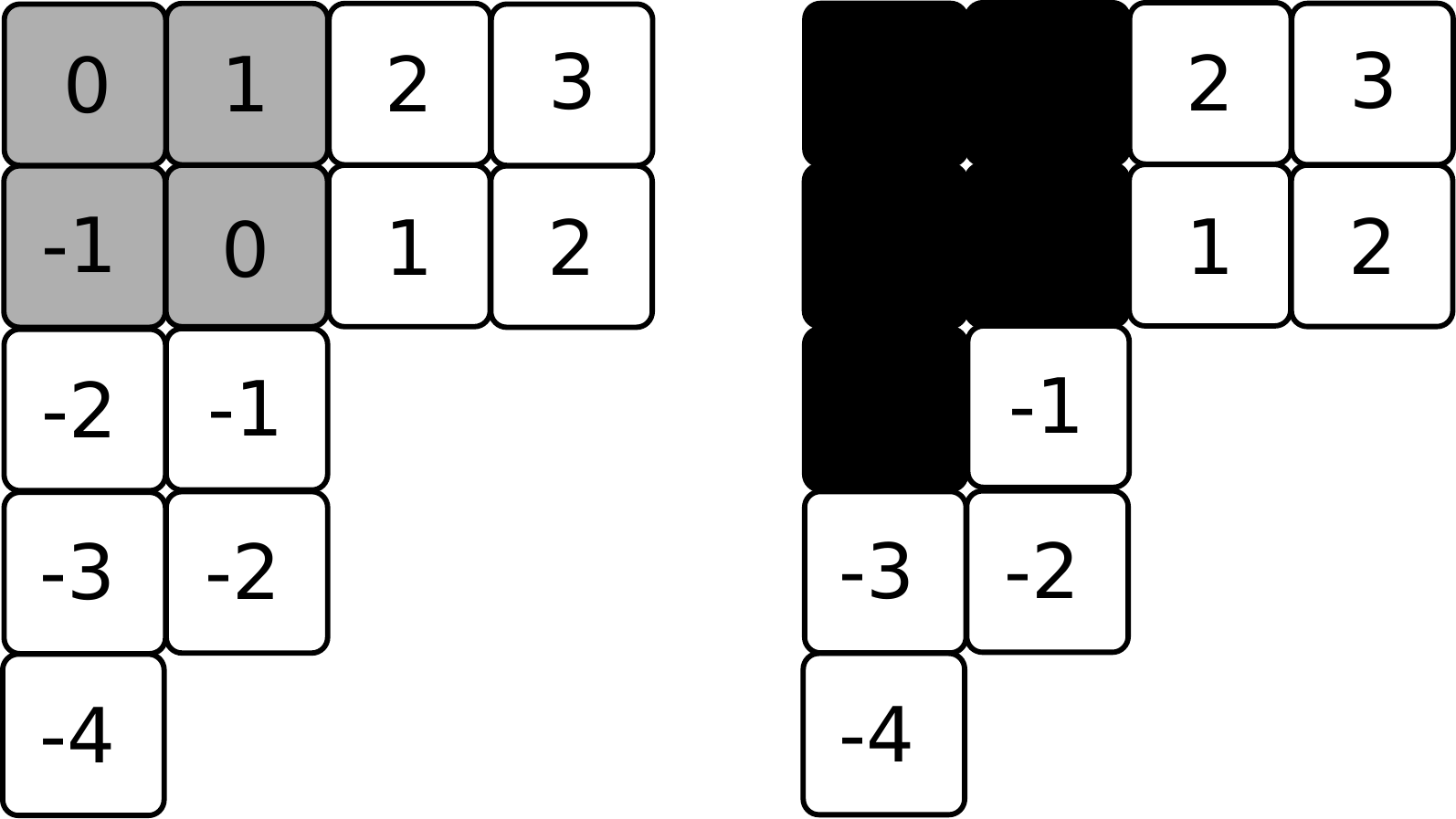}
\caption{Left: Diagram of the partition $(4,4,2,2,1)$, showing contents; the Durfee square is highlighted in grey. Right: the skew diagram $(4,4,2,2,1)/(2,2,1)$ and its contents.} 
\end{figure}

When $\lambda\supset\mu$, the skew shape $\lambda/\mu$ exists and is represented by the diagram of $\lambda$ with the boxes contained in the diagram of $\mu$ being removed. An example is shown Figure 5 in which $\lambda=(4,4,2,2,1)$ and $\mu=(2,2,1)$. In this case $t_\mu=2$ and $t_{\lambda/\mu}=t_\lambda/t_\mu=288$.

\section*{Appendix B}

We want to expand 
\be s_\mu\left(\frac{X}{1-\gamma X}\right)=\sum_\rho A_{\mu\rho}s_\rho(X). \ee The key to this expansion is the fact that Schur functions, being irreducible characters of the unitary group, satisfy the Weyl orthogonality relation
\be \int_{U(N)} s_\lambda(U)s_\mu(U^\dagger)dU=\frac{1}{N!(2\pi)^N}\oint s_\lambda(z)s_\mu(\bar{z})|\Delta(z)|^2,\ee where there are $N$ variables $z$, all being integrated around the unit circle. Here $\bar{z}$ is the complex conjugate. From the orthogonality relation it follows immediately that
\be A_{\mu\rho}=\frac{1}{N!(2\pi)^N}\oint  s_\mu\left(\frac{z}{1-\gamma z}\right)s_\rho(\bar{z})|\Delta(z)|^2.\ee

Using the expression of the Schur functions as a determinant we can write
\be s_\mu\left(\frac{z}{1-\gamma z}\right)=\frac{1}{\Delta\left(\frac{z}{1-\gamma z}\right)}{\rm det}\left[\left(\frac{z_k}{1-\gamma z_k}\right)^{N+\mu_i-i}\right].\ee The Vandermonde in the denominador is \be \Delta\left(\frac{z}{1-\gamma z}\right)=\frac{\Delta(z)}{\prod_k(1-\gamma z_k)^{N-1}}.\ee Therefore,
\be A_{\mu\rho}=\frac{1}{N!(2\pi)^N}\oint  {\rm det}\left(\frac{z_k^{N+\mu_i-i}}{(1-\gamma z_k)^{\mu_i-i+1}}\right){\rm det}(\bar{z}_k^{N+\rho_j-j}).\ee

Using the Andreief lemma, this becomes
\be A_{\mu\rho}=\frac{1}{(2\pi)^N}{\rm det}\left[\oint \frac{z^{N+\mu_i-i}}{(1-\gamma z)^{\mu_i-i+1}}\bar{z}^{N+\rho_j-j}\right].\ee

Since $z^N\bar{z}^N=1$, the binomial theorem leads to
\be A_{\mu\rho}=\frac{1}{(2\pi)^N}{\rm det}\left[ \sum_{k=0}^\infty {\mu_i-i+k\choose \mu_i-i}\gamma^k \oint z^{\mu_i-i+k}\bar{z}^{\rho_j-j}\right].\ee

Now the crucial point is that
\be \oint z^a\bar{z}^b=2\pi\delta_{ab}.\ee So the integral is only different from zero if $k=\rho_j-j-\mu_i+i$. Since $k\ge 0$ we have that $\rho_j-j\ge\mu_i-i$ and
\be A_{\mu\rho}={\rm det}\left[ {\rho_j-j\choose \mu_i-i}\gamma^{\rho_j-j-\mu_i+i}\right].\ee
Expanding the determinant we get $\prod_j \gamma^{\mu_j}=\gamma^{|\mu|}$ and likewise for $\rho$. In the end,
\be A_{\mu\rho}=\gamma^{|\rho|-|\mu|}{\rm det}\left[ {\rho_j-j\choose \mu_i-i}\right].\ee

\end{document}